\documentclass{ws-mpla}
\usepackage{slashed}

\begin{document}

\markboth{Alexander Rothkopf}
{From Complex to Stochastic Potential}

%%%%%%%%%%%%%%%%%%%%% Publisher's Area please ignore %%%%%%%%%%%%%%
\catchline{}{}{}{}{}
%%%%%%%%%%%%%%%%%%%%%%%%%%%%%%%%%%%%%%%%%%%%%%%%%%%%%%%%%%%%%%%%%%%

\title{From Complex to Stochastic Potential:\\
Heavy Quarkonia in the Quark-Gluon Plasma
}

\author{\footnotesize Alexander Rothkopf}

\address{Albert Einstein Center for Fundamental Physics, University of Bern, Sidlerstrasse 5,
Bern, 3012,
Switzerland\\
rothkopf@itp.unibe.ch}

\maketitle

\begin{abstract}
The in-medium physics of heavy quarkonium is an ideal proving ground for our ability to connect knowledge about the fundamental laws of physics to phenomenological predictions. One possible route to take is to attempt a description of heavy quark bound states at finite temperature through a Schr\"odinger equation with an instantaneous potential. Here we review recent progress in devising a comprehensive approach to \textit{define} such a potential from first principles QCD and \textit{extract} its, in general complex, values from non-perturbative lattice QCD simulations. Based on the theory of open quantum systems we will show how to \textit{interpret} the role of the imaginary part in terms of spatial decoherence by introducing the concept of a stochastic potential. Shortcomings as well as possible paths for improvement are discussed.

\end{abstract}

%\ccode{PACS Nos.: include PACS Nos.}

\section{Motivation}
The investigation of the quark gluon plasma created in relativistic heavy-ion collisions relies on a fruitful interplay between theory and experiment. Hence the availability of observables that are both experimentally accessible as well as theoretically amenable is of importance. The first unambiguous sign for the creation of a deconfined and strongly interacting state of matter, reported by the RHIC experiments PHENIX\cite{Adcox:2004mh} and STAR\cite{Adams:2005dq}, e.g. relied on the occurrence of collective flow\cite{Afanasiev:2009wq,Adamczyk:2011aa} and the quenching of jets \cite{Adler:2005ee,Adler:2002tq} in $\rm Au+Au$ collisions at $\sqrt{s_{\rm NN}}=200 {\rm GeV}$.

Early on, Matsui and Satz\cite{Matsui:1986dk} proposed another type of hard probe, the bound states of heavy quarks. From analogies with the electromagnetic plasma, they concluded that a Debye screened potential above the deconfinement temperature would not be able to support the formation of a $c\bar{c}$ bound state. This in turn would lead to a suppression of the measured abundance of heavy quarkonia in the presence of a quark-gluon plasma. 

While heavy quarkonium suppression has indeed been confirmed\cite{Adare:2006ns,Tang:2011kr,Chatrchyan:2012np,Abelev:2012rv} by several experiments, it turns out, that the physics of charmonium in relativistic heavy-ion collisions is rather involved. Cold nuclear matter effects, as well as final state effects, such as recombination and feed-down mask the actual influence of the thermal medium on the stability of the bound state. 

In the case of $b\bar{b}$, which became measurable with adequate statistics, once the LHC commenced its runs at $\sqrt{s_{\rm NN}}=2.76{\rm TeV}$, the connection is much cleaner. Produced in the hard partonic stage at the beginning of the collision, bottomonium does not suffer from significant recombination effects nor is there appreciable feed-down from $t\bar{t}$. The dimuon spectra measured by the CMS collaboration\cite{Chatrchyan:2011pe} furthermore provided clear evidence for the suppression of the excited states in $\rm Pb+Pb$ collisions compared to their abundances in $\rm p+p$. Under these conditions heavy quarkonium becomes an ideal candidate to investigate the physics of melting bound states, which will be our focus in the rest of this brief review.

Ever since Matsui and Satz, the goal for theory has to be to put their phenomenological arguments onto a solid first-principles basis. In the past many attempts relied on a time independent notion of the melting process, introducing purely real model potentials and investigating their ability to support a bound state of $Q$ and $\bar{Q}$. Popular candidates were the color singlet free energies $F^1(r)$\cite{Nadkarni:1986as}, the color singlet internal energies $U^1(r)$\cite{Kaczmarek:1900zz} and even linear combinations of both quantities\cite{Satz:2005hx}. While these quantities can be readily extracted from lattice QCD correlation functions at a single time step $\beta=\tau$, a direct connection to a Schroedinger equation derived from first principles QCD is unlikely to exist\cite{Burnier:2009bk}.

In a situation where the heavy quarkonium is produced inside the quark-gluon plasma, e.g. through recombination, questions about the possibility of bound state formation can lead to insight. Bottomonium however is created before the QGP thermalizes and we need to understand its real-time evolution inside the medium to eventually grasp the physics of its suppression.

An essential step into this direction was the discovery by Laine et.al.\cite{Laine:2006ns} that the heavy quark potential at first non-trivial order in resummed perturbation theory not only shows Debye screening but also features an imaginary part. Its appearance was subsequently attributed to the phenomenon of Landau damping\cite{Beraudo:2007ky}. First corrections to these results in the context of effective field theory were presented in\cite{Brambilla:2008cx} introducing further contributions to both real and imaginary part. These findings remind us that a static notion of a well defined bound state above the deconfinement temperature becomes devoid of meaning and urge us to devise a fully dynamic approach to the stability and melting of heavy quarkonium instead.

Let us begin by asking how a non-relativistic Schroedinger equation with instantaneous potential can be consistently derived from the underlying field theory of QCD.

\vspace{-0.5cm}
\section{Defining the static heavy quark potential}

What makes heavy quarkonia particularly suitable to theoretical treatment is the inherent separation of scales between its constituent mass\cite{Beringer:1900zz} $(m_c\simeq1.29{\rm GeV},m_b\simeq4.6{\rm GeV})$ and other typical scales in relativistic heavy ion-collisions. Compared to e.g. the temperature of the deconfinement transition $T_C\sim200{\rm MeV}$, the intrinsic scale of QCD $\Lambda_{\rm QCD}\sim200{\rm MeV}$ and the typical momentum exchange in the bound state ${\bf p}$, we find \vspace{-0.3cm}
\begin{align}
 \frac{T_C}{m_Q} \ll 1,\quad \frac{\Lambda_{\rm QCD}}{m_Q}\ll 1, \quad \frac{\bm p}{m_Q}\ll1.
\end{align}
This in turn tells us that neither thermal nor quantum fluctuations can spontaneously pair create a heavy $Q$ and $\bar{Q}$ and thus a non-relativistic description can be attempted.

Whenever there exists a hierarchy of scales, effective field theory (EFT) provides a general framework on how to reduce the complexity of the description by treating explicitly only the physics at the energy scales of interest. (For a review of the technique applied to QCD at T=0 see Refs.\cite{Brambilla:1999xf,Brambilla:2004jw} and references therein.) In our case the physics of the bound state is well separated from the hard scale $m_Q$, which we hence wish to integrate out. The notion of ``integrating out`` encompasses three actions: First we need to identify the relevant degrees of freedom at the energy scale of interest; secondly the most general Lagrangian in these new variables is to be constructed while retaining the symmetries of the underlying field theory. Last but not least, the residual influence of the higher lying energy scale is incorporated through the process of matching, where suitable correlation functions in the original and effective language are made to agree at a certain energy scale.

\subsection{An effective field theory of Pauli spinors}
\label{Sec:DiracPauli}
The first step on the path towards a heavy quark potential thus lies in applying the above concepts to QCD. We begin by integrating out the hard scale, which leads us to the effective field theory NRQCD. The starting point is the QCD Lagrangian
\begin{align}
\nonumber &{\cal L}_{\rm QCD} = \\
\nonumber &-\frac{1}{4} F^{\mu\nu}_aF_{\mu\nu}^a + \sum_{l=1}^{N_f}\bar{q}^l\Big(i\gamma^\mu (\partial_\mu + igA_\mu) -m^l_q\Big) q^l + \bar{Q} \Big(i\gamma^\mu (\partial_\mu + igA_\mu) -m_Q\Big) Q, %\label{QCDLag}
\end{align}
written in terms of matrix valued gauge fields $A^\mu=A^\mu_aT^a$, light medium quarks $q$ and heavy quark fields denoted by $Q$.

As the physics below the hard scale does not involve pair creation of heavy quarks, the upper and lower components of the Dirac spinor $Q=(\xi,\chi)$ themselves suffice to capture the relevant physics and we proceed to separate $\xi$ and $\chi$ by means of a Foldy-Tani-Wouthuysen (FTW) transformation\cite{Tani:1951aa,Foldy:1949wa}. The resulting NRQCD Lagrangian presents itself as a series of terms of increasing powers in the inverse heavy quark mass $m_Q^{-1}$, the first few of which read\cite{Brambilla:2004jw}
\begin{align}
 \nonumber{\cal L}_{\rm NRQCD} = &-\frac{1}{4} F^{\mu\nu}_aF_{\mu\nu}^a + \sum_{l=1}^{N_f}\bar{q}^l (\slashed{D}-m_q) q^l+\xi^\dagger(iD_0-m_Q+\frac{c_1}{2m_Q}{\bf D}^2+\ldots)\xi\\
 &+ \chi^\dagger(iD_0-m_Q+\frac{c_1}{2m_Q}{\bf D}^2+\ldots)^\dagger\chi\label{NRQCDLag}.
\end{align}
Note that in addition to the transformation of fields, we have inserted so called Wilson coefficients $c_i$ in each term that scales explicitly with a power of $m_Q^{-1}$. These complex numbers encode the remnants of the physics at higher energy scales, such as e.g. a possible  contribution of a gluonic cloud to a kinetic mass of the quarks via $c_1$. One difference to the usual NRQCD approach is that here we do not integrate out gluons and light quarks with hard momenta, as we assumed $T_{\rm med}\ll m_Q$ and our aim is to explicitly incorporate their contribution through the use of non-perturbative lattice QCD simulations. For our purpose of deriving a static and spin-independent potential, we can safely set all values of $c_i$ to unity in the following. (For further information on the explicit determination of the $c_i$'s through matching we refer the reader to Ref.\cite{Brambilla:2004jw}).

The strength of the FTW transformation is to yield a systematic expansion in the inverse rest mass, which at the same time alerts us to the limits of any potential picture derived via NRQCD. Since after summing up an infinite number of terms, the underlying QCD Lagrangian emerges, couplings between the upper and lower components $\chi$ and $\xi$ will necessarily reappear at some order and the absence of explicit pair creation is not guaranteed. The first relevant terms, such as $\xi^\dagger\chi\chi^\dagger\xi$ indeed appear at order $m_Q^{-2}$, signaling the break down of the non-relativistic approximation at higher orders.

The notion of potential has so far not found entry into the description of the dynamics, as the Pauli spinor fields $\xi$ and $\chi$ interact via the explicit mediation of the gauge fields $A^\mu$. One possibility to introduce $V(r)$ is to integrate out all degrees of freedom down to the so called soft scale, characterized by energies $E\sim m_Q v$. The Wilson coefficients in this approach are non-local and represent potentials between well defined color singlet and octet configurations of the heavy quarks. Termed pNRQCD, this effective field theory allows a consistent power counting and the choice of degrees of freedom in the form of correlated pairs of heavy quarks has proven successful at zero temperature\cite{Brambilla:1999xf} as well as in in-medium applications\cite{Brambilla:2008cx}.

On the other hand, in the quark-gluon plasma we expect the heavy quark bound states to melt eventually and thus ultimately our description should comprise as relevant degrees of freedom also decorrelated pairs of individual heavy quarks.

\subsection{Quantum mechanical path integrals}
\label{Sec:QMPI}
The question remains how to describe the heavy quark bound state while emphasizing the propagation of its individual constituents? One possible way is to turn to non-relativistic path integrals\cite{Barchielli:1986zs,Barchielli:1988zp}, which operate with point particles\cite{Feynman:1948ur} along fluctuating trajectories ${\bf z}_i$ and their conjugate momenta ${\bf p}_i$. In particular they will allow us to to read off the interaction potential by applying the transfer matrix prescription to the complex weighting factor ${\rm exp}[i\int dt({\bf p}\dot{{\bf z}}-H)]$ of a two-body path integral. (For an application to a single heavy quark see Ref.\cite{Beraudo:2010tw})

Starting point is a suitable correlation function in field theory that serves as propagator for the heavy quarkonium system. Here we choose the forward correlator $D^>$, as it encodes how probable it is to find a bare $Q\bar{Q}$ state at time t after starting the evolution at time t=0
\begin{align}
 D^>({\mathbf x}_1,{\mathbf y}_1,&{\mathbf x}_2,{\mathbf y}_2,t) =\label{Eq:ForwCorr} \\
\nonumber & \langle \bar{Q}({\mathbf x}_2,t)U({\mathbf x}_2,{\mathbf y}_2,t)Q({\mathbf y}_2,t)  \bar{Q}({\mathbf y}_1,0)U^\dagger({\mathbf x}_1,{\mathbf y}_1,0)Q({\mathbf x}_1,0)\rangle
\end{align}
For the expression to be gauge invariant we connected the point split fermionic fields by straight spatial Wilson lines $U( {\bf x}, {\bf y},t)$. 

The results of Sec.\ref{Sec:DiracPauli} allow us to to reexpress Eq.\eqref{Eq:ForwCorr} in the Pauli spinor fields of the NRQCD effective field theory
\begin{align}
 \nonumber &D^>_{\rm NRQCD}({\mathbf x}_1,{\mathbf y}_1,{\mathbf x}_2,{\mathbf y}_2,t)=\\
 &\int D[\xi,\bar{\xi}] D[\chi,\bar{\chi}] D[A,q,\bar{q}] \xi^\dagger({\mathbf x}_2,t)U\chi({\mathbf y}_2,t)  \chi^\dagger({\mathbf y}_1,0)U^\dagger\xi({\mathbf x}_1,0) e^{iS_{\rm NRQCD}}.    
\end{align}
The trick here is to actually carry out analytically the quadratic integration over the heavy Gra\ss mann fields, which leaves us with non relativistic propagators, describing the evolution of quark and anti-quark in the background of the medium gluon field $A^\mu$. As heavy quarks in NRQCD, the fields $\xi$ and $\chi$ cannot participate in virtual loops and thus no fermion determinant has to be taken care of. In practice we have to replace pairs of $\xi\xi^\dagger$ and $\chi\chi^\dagger$ by the Green's functions $K$ and $K^\dagger$, i.e. the functional inverses of the NRQCD action defined by
\begin{align}
\nonumber  \Big[i\partial_t - g&A_0 - m_Q +\frac{1}{2m}\Big(\partial_j+\frac{ig}{c}A_j\Big)^2\Big]K({\mathbf x}_1,{\mathbf x}_2)=0,\\
\nonumber &\lim_{{\mathbf x}_1\to {\mathbf x}_2} K({\mathbf x}_1,{\mathbf x}_2)=\delta^{(3)}({\mathbf x}_1-{\mathbf x}_2).
\end{align} 
It is at this point that the connection to quantum mechanical path integrals is made, as the propagator of a particle with trajectory $\bf z$ and momentum $\bf p$ can be expressed as
\begin{align}
\nonumber K({\mathbf x}_1,{\mathbf x}_2)=\int_{{\mathbf x}_1}^{{\mathbf x}_2} {\cal D}{\bf z}\int {\cal D}{\bf p}  \; {\cal T} \; {\rm exp}\Bigg[i\int_{0}^{t} ds \Big( {\bf p}\dot{{\bf z}} + \frac{{\bf p}^2}{2m_Q}-m_Q\Big)\Bigg] {\rm exp}\Bigg[ig\int_{\mathbf z} dy^\mu A_\mu(y) \Bigg].%\label{Eq:PathIntK}
\end{align}
Since quark and anti-quark each enter through a separate path integration, the forward propagator in the language of quantum mechanics takes the form
\begin{align}
 D^>_{\rm QM}=&e^{-2im_Qt} \int_{{\mathbf x}_1}^{{\mathbf x}_2} {\cal D}[{\bf z}_1,{\bf p}_1]\int_{{\mathbf y}_1}^{{\mathbf y}_2} {\cal D}[{\bf z}_2,{\bf p}_2]\; \label{Eq:FullQMPathInt} \\
\nonumber& \times{\rm exp}\Big[ i\sum_{l=1}^2\int_0^{t} ds \Big(  {\bf p}_l(s) \dot{{\bf z}}_l(s) + \frac{{\bf p}_l^2(s)}{2m}\Big)\Big] \left\langle {\cal T} {\rm exp} \Big[  \frac{ig}{c} \oint dx_\mu A_\mu(x)\Big] \right\rangle,
\end{align}
where we have separated a field independent term containing the kinetic terms, quadratic in the momenta, and a field dependent term, which still requires us to average over the medium degrees of freedom. 

In order to understand how a potential of the two-body system emerges from Eq.\eqref{Eq:FullQMPathInt}, we focus on its last term, which is nothing but the thermal real-time Wilson loop\footnote{Even though this quantity is usually referred to as thermal, it is not periodic in imaginary time. Intuitively this is a result of the static quarks not being thermalized. The gluons of the thermal medium and their periodicity however introduce an upward trend into the Euclidean time data, seen most clearly as $\tau\to\beta$.}. If a potential picture were applicable at all times, we expect to be able to rewrite the expression as an exponential over a time independent function $V(r=|{\bf z}_2-{\bf z}_1|)$. Since what we do however amounts to replacing a retarded field theoretical interaction between quark and anti-quark by an instantaneous non-relativistic potential, we have to take into account the possibility that the function $V(r,s)$ does actually vary at early times before approaching a constant value only 
at late times \vspace{-0.2cm}
\begin{align}
  W_\square(r,t)=\left\langle {\cal T} {\rm exp} \Big[  ig \int_\square dx_\mu A_\mu(x)\Big] \right\rangle = {\rm exp}\Big[-i\int_0^t \; ds\; V(r,s)\Big]\label{Eq:WLoopPot}.
\end{align}
Note that we assume that the heavy mass limit allows us to neglect terms connecting different times along the path of the quark-antiquark pair in Eq.\eqref{Eq:WLoopPot}. Going to late times finally leads us to the following defining equation, connecting the time evolution of the real-time Wilson loop to the static inter-quark potential
\begin{align}
 V(r)=\lim_{t\to\infty}\frac{i\partial_t W_\square(r,t)}{W_\square(r,t)}\label{Eq:DefPot}.
\end{align}

\section{Extracting the static heavy quark potential}

Now that the field theory of QCD in form of the real-time Wilson loop has been connected to a non-relativistic potential description of heavy quarkonia in Eq.\eqref{Eq:DefPot}, we have to ask how the actual values of such a  potential can be determined in practice. 

One possibility, worked out in detail in the ground-breaking contribution by Laine et. al.\cite{Laine:2006ns} is to use resummed perturbation theory, the so called hard thermal loop approximation\cite{Pisarski:1988vd,Braaten:1989mz}, in describing the medium the heavy quarks travel in. This gauge invariant prescription allows us to sum an infinite number of Feynman diagrams for thermal gluons and light quarks and already captures essential features of the in-medium physics at high temperatures. Evaluating the Wilson loop at first non-trivial order in the gauge coupling and inserting into Eq.\eqref{Eq:DefPot} yields\cite{Laine:2006ns}
\begin{align}
 V_{\rm HTL}(r)&=-\frac{g}{3\pi}\Big[m_D+\frac{e^{-m_D r}}{r}\Big] - \frac{ig^2T}{3\pi} \phi(m_Dr), \label{Eq:LainePot}\\
 \nonumber \phi(x)&=2\int_0^\infty dz \frac{z}{(z^2+1)^2}\Big[1-\frac{sin[zx]}{zx}\Big].
\end{align}
One finds that the potential is complex with the real part exhibiting a Debye screened form with screening mass $m_D^2=g^2T^2\Big(\frac{N_c}{3}+\frac{N_f}{6}\Big)$. The appearance of an imaginary part on the other hand can be linked to collisions between the light partons in the medium and the gluon mediating the interaction between the $Q$ and $\bar{Q}$, i.e. the phenomenon of Landau damping\cite{Beraudo:2007ky}. While ${\rm Im}[V_{\rm HTL}]$ grows quadratically at small $m_Dr<1$ it saturates to a constant value at large $m_Dr\gg1$, which in turn can be interpreted as twice the energy loss of a single quark traversing a heat bath.
First corrections to Eq.\eqref{Eq:LainePot} have been presented in a pNRQCD context\cite{Brambilla:2008cx}, incorporating the first oder of the multipole expansion. There it was shown that the breakup of a singlet to an octet state under the influence of a color electric field can also contribute to the imaginary part.

While asymptotic freedom guarantees the correctness of weak-coupling results at high temperatures, we expect the quark-gluon plasma to be strongly interacting in the phenomenologically important region around the phase transition\cite{Sarkar:2010zza}. One novel possibility to, at least qualitatively, elucidate such non-perturbative QGP physics is the gauge-gravity duality\cite{Maldacena:1997re}. By mapping a certain type of strongly coupled conformal Yang-Mills theory to a weakly coupled classical field theory in higher dimensions, calculations that are technically impossible in the former can actually be carried out in the latter. This approach has recently been used\cite{Hayata:2012rw} to determine the values of the thermal real-time Wilson loop and subsequently the static potential according to the definition Eq.\eqref{Eq:DefPot}. While the obtained real part at small distances shows Coulombic behavior, at large distances it exhibits a rather long tail, whose value decreases with increasing temperature. The 
imaginary part on the other hand while being zero up to a threshold $r_{\rm th}=0.62(\pi T)^{-1}$ runs linearly in distance already dominating the real part at relatively short $r_{\rm dom}=1.72(\pi T)^{-1}$.

In the following we investigate the non-perturbative aspects of heavy-quarkonium physics, while staying firmly within the framework of QCD. One possible way to do so is to rely on Monte Carlo simulations of spatially regularized, so called lattice QCD. This approach allows us to evaluate observables at any temperature, with the only but significant limitation that all calculations have to be carried out in imaginary time instead of real-time. Hence no direct access to the real-time Wilson loop of Eq.\eqref{Eq:DefPot} is possible. In order to circumvent this obstacle and to obtain the sought after information necessary to evaluate the potential, we will introduce a spectral representation of the Wilson loop instead\cite{Rothkopf:2009pk,Rothkopf:2011db}.

\subsection{A spectral representation}

It has been shown\cite{Rothkopf:2009pk} that the Wilson loop admits a spectral representation, which amounts to nothing but a Fourier transform over a positive definite function $\rho_\square(r,\omega)$
\vspace{-0.5cm}
\begin{align}
 W_\square(r,t)=\int_{-\infty}^\infty d\omega e^{-i\omega t} \rho_\square(r,\omega) \label{Eq:WilsonLoopFour}.
\end{align}
The benefit of writing the Wilson loop in this way is that its time dependence only appears in the kernel of the integral. If the spectral information is known in the form of $\rho_\square(r,\omega)$, changing from real- to imaginary time just means to go over from a Fourier to a Laplace transformation \vspace{-0.1cm}
\begin{align}
 W_\square(r,\tau)=\int_{-\infty}^\infty d\omega e^{-\omega \tau} \rho_\square(r,\omega)\label{Eq:WilsonLoopLap}.
\end{align}
Technically there is however an important difference, since an exponentially damping Laplace kernel makes the inversion of the relation between spectrum and Wilson loop much more demanding than an oscillating Fourier kernel. And it is exactly such an inversion that we have to perform if we wish to extract from the noisy lattice QCD estimates of $W_\square(r,\tau)$ the spectral information to reconstruct $W_\square(r,t)$\footnote{Note that even though the Euclidean time quantities are all purely real, the presence of the Fourier kernel makes it possible for the potential to have an imaginary part in the end.} (See Sec. \ref{Sec:MaxEntr}). 

The question to ask here is whether we can already extract the values of the potential from spectral information of the Wilson loop? To find out we follow\cite{Rothkopf:2011db} by inserting the spectral representation of Eq.\eqref{Eq:WilsonLoopFour} into the defining Eq.\eqref{Eq:DefPot} and obtain
\begin{align}
 V(r)=\lim_{t\to\infty}\frac{ \int_{-\infty}^\infty d\omega\; \omega\; e^{-i\omega t} \;\rho_\square(r,\omega)}{\int_{-\infty}^\infty d\omega \; e^{-i\omega t} \; \rho_\square(r,\omega)}\label{Eq:DefSpecPot}.
\end{align}
Even though we know from Eq.\eqref{Eq:DefPot} that the potential is only connected to late time physics and naively speaking Eq.\eqref{Eq:DefSpecPot} hence signals that we will need to consider spectral features only at small frequencies, the fact that the Fourier transform relies on late and early time information needs to be kept in mind.

Indeed from our derivation of the potential in Sec.\ref{Sec:QMPI} we know\cite{Burnier:2012az} that the description of the time evolution of the Wilson loop in the form
\begin{align}
 i\partial_tW_\square(r,t)=\Phi(r,t)W_\square(r,t)\label{Eq:WilsonLoopTimeEvol}
\end{align}
in general requires a complex function $\Phi(r,t)$. Only at late times it approaches a constant, which we identify with the potential
\begin{align}
 \lim_{t\to\infty}\Phi(r,t)=V(r)\label{Eq:PotAsympt}.
\end{align}

\vspace{-0.1cm}

Following Ref.\cite{Burnier:2012az} we introduce an additional function $\phi(r,t)$, which encodes the deviation of $\Phi(r,t)=V(r)+\phi(r,t)$ from the potential and solve Eq.\eqref{Eq:WilsonLoopTimeEvol} to obtain a general expression for the Wilson loop \vspace{-0.1cm}
\begin{align}
 \nonumber W_\square(r,t)={\rm exp}\Bigg[ -i\Big(& {\rm Re}[V](r)t + {\rm Re}[\sigma](r,t)\Big) -|{\rm Im}[V](r)|t+{\rm Im}[\sigma](r,t) \Bigg]\label{Eq:WLPotMod}.
\end{align}
The quantity $\sigma(r,t)=\int_0^t\phi(r,t)dt$ and its asymptotic values are defined as $\sigma_\infty(r)=\sigma(r,|t|>t_{Q\bar{Q}})=\int_0^\infty\phi(r,t)dt$. Carrying out the Fourier transform\cite{Burnier:2012az} tells us that the most general spectral shape we will encounter at low frequencies in $\rho_\square(r,\omega)$ reads
 \begin{eqnarray}
 \nonumber \rho_\square(r,&&\omega)=\frac{1}{\pi}e^{{\rm Im}[\sigma_\infty](r)} \frac{|{\rm Im}[V](r)|{\rm cos}[{\rm Re}[\sigma_\infty](r)]-({\rm Re}[V](r)-\omega){\rm sin}[{\rm Re}[\sigma_\infty](r)]}{ {\rm Im}[V](r)^2+ ({\rm Re}[V](r)-\omega)^2}\\&&+\kappa_0(r)+\kappa_1(r)t_{Q\bar Q}({\rm Re}[V](r)-\omega)+\kappa_2(r)t_{Q\bar Q}^2({\rm Re}[V](r)-\omega)^2+\cdots\label{Eq:FitShapeFull}
\end{eqnarray}
The first term in the above Eq.\eqref{Eq:FitShapeFull} takes the form of a skewed Lorentzian embedded in a polynomial background, characterized by real numbers $\kappa_i(r)$. Even though the result in Eq.\eqref{Eq:FitShapeFull} appears quite involved, inserting it into our defining formula Eq.\eqref{Eq:DefSpecPot} and carrying out the contour integration in the lower half of the complex plane yields a time independent potential with $V(r)={\rm Re}[V](r)-i|{\rm Im}[V](r)|$. In earlier studies\cite{Rothkopf:2011db} the spectral features were assumed to be a perfect Breit-Wigner, which we can now understand from the above derivation as corresponding to the additional assumption $\partial_t\Phi(r,t)=0$ , i.e. that the potential picture be applicable at all times.

\subsection{The maximum entropy method}
\label{Sec:MaxEntr}
All conceptual ingredients are now in place to connect the spectral information of the Wilson loop to the concept of a static inter-quark potential. One remaining issue of practical importance is how the spectrum is actually measured in lattice QCD. Over the last decade a Bayesian approach to this question has found widespread adoption, the Maximum Entropy Method\cite{Nakahara:1999vy,Asakawa:2000tr} (MEM). What we are facing in the context of lattice QCD when attempting to invert Eq.\eqref{Eq:WilsonLoopLap} is a inherently ill-defined problem. The reason is that the Monte-Carlo estimates of $W_\square(r,\tau)$ are available only on a discrete number of points and carry a finite uncertainty. If we were to attempt an extraction of a continuous function $\rho_\square(r,\omega)$ via $\chi^2$ fitting, we immediately find that an infinite number of degenerate solutions will ensue, all of which fit the data within their errorbars.

Based on Bayes theorem, we can nevertheless give meaning to the inversion task by emphasizing the role of prior information $I$. The probability of a test spectral function to be the correct spectral function, given measured data $D$ and prior information $I$, can be written as the product of two terms 
\begin{align}
P[\rho|D,I]\propto P[D|\rho]P[\rho|I].
\end{align}
The first one denotes the so called Likelihood probability, the second one refers to the prior probability. While $P[D|\rho]$ is nothing but the usual $\chi^2$ fitting term, the MEM uses the so called Shannon-Jaynes entropy ${\cal S}$ in $P[\rho|I]$ to incorporate prior information, in the form of a function $m(\omega)$, while enforcing the positive definiteness of $\rho$ \vspace{-0.3cm}
\begin{align}
 P_{MEM}[\rho|I(m)]\propto{\rm exp}[\alpha {\cal S}] = {\rm exp}\Big[ \alpha \sum_{l=1}^{N_\omega} \Big( \rho_l-m_l-\rho_l{\rm log}[\frac{\rho_l}{m_l}]\Big)\Big]. \label{Eq:PriorProb}
\end{align}
If we set out to determine the most probable spectral function, we are lead to numerically solve the following stationarity condition
\begin{align}
 \left. \frac{\delta}{\delta \rho_l} P[\rho|D,I(m)] \right|_{\rho=\rho_{\rm MEM}}\propto \left. \frac{\delta}{\delta \rho_l}\Big( P[D|\rho]P_{MEM}[\rho,I]\Big) \right|_{\rho=\rho_{\rm MEM}}=    0.\label{MEM:Optimize}
\end{align}
Here the otherwise underdetermined $\chi^2$ fitting, which attempts to solely reproduce the measured data, competes with the prior term which tries to bring the spectrum as close to the prior function $m(\omega)$ as possible. It can be shown\cite{Asakawa:2000tr} that Eq.\eqref{MEM:Optimize} actually possesses a unique solution $\rho_{\rm MEM}$ if such a solution exists. Intuitively the reason is that by taking together measured datapoints and the supplied prior function, we have at our disposal more points of information than parameters to extract. Note that in the extreme case of no measured datapoints, by definition, the function $m(\omega)$ itself represents the correct spectrum.

Consequently, parts of the spectrum $\rho_{\rm MEM}$ are constrained by the measured data, parts of it by the provided prior function. The introduction of such a regularizing mechanism into the $\chi^2$ fitting hence allows us to unambiguously identify what spectral features are reliably encoded in the measured data, i.e. by repeating the MEM extraction with many different functional forms of the prior $m(\omega)$. For technical details on the numerical implementation and the question of an appropriate search space in which the extremum, defined by Eq.\eqref{MEM:Optimize}, can be located, we refer the reader to Refs.\cite{Asakawa:2000tr,Jakovac:2006sf,Rothkopf:2011ef,Rothkopf:2012vv}.

\begin{figure}[t!]
\centerline{\includegraphics[scale=0.55,clip=true, trim=0 7.5cm 0 5.1cm]{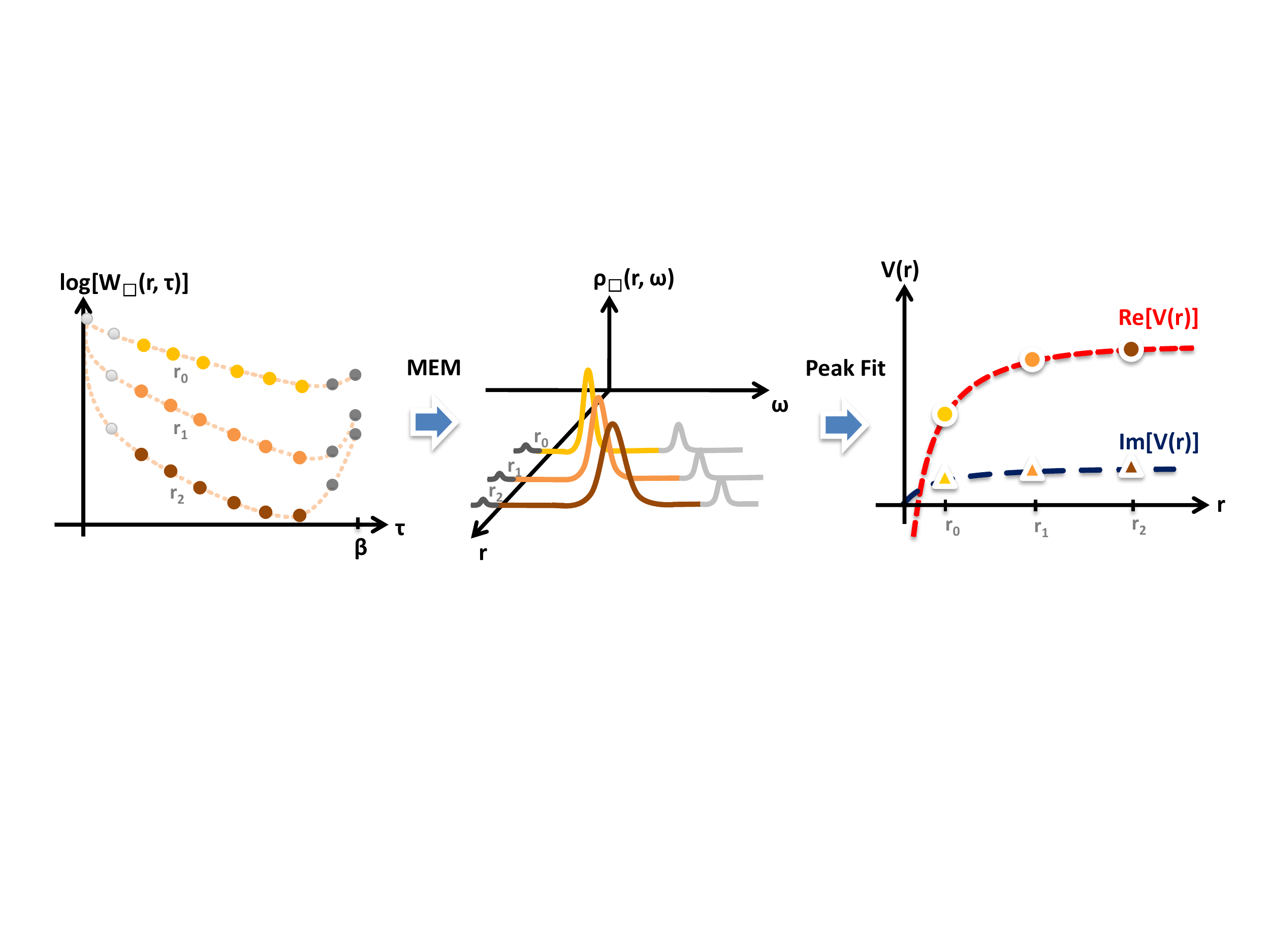}}
\vspace*{8pt}
\caption{A schematic overview of the strategy to non-perturbatively extract the static heavy-quark potential from lattice QCD simulations. From measurements of the Euclidean Wilson loop at different separation distances along the imaginary time axis, we can obtain the corresponding spectral information by applying the Maximum Entropy Method. Based on the most general functional form given in Eq.\eqref{Eq:FitShapeFull} the lowest lying spectral feature at positive frequencies is fitted and the values of ${\rm Re}[V](r)$ and ${\rm Im}[V](r)$ are read off. \protect\label{fig1}}
\end{figure}

\begin{figure}[t!]
\centerline{\includegraphics[scale=0.2,angle=-90,clip=true]{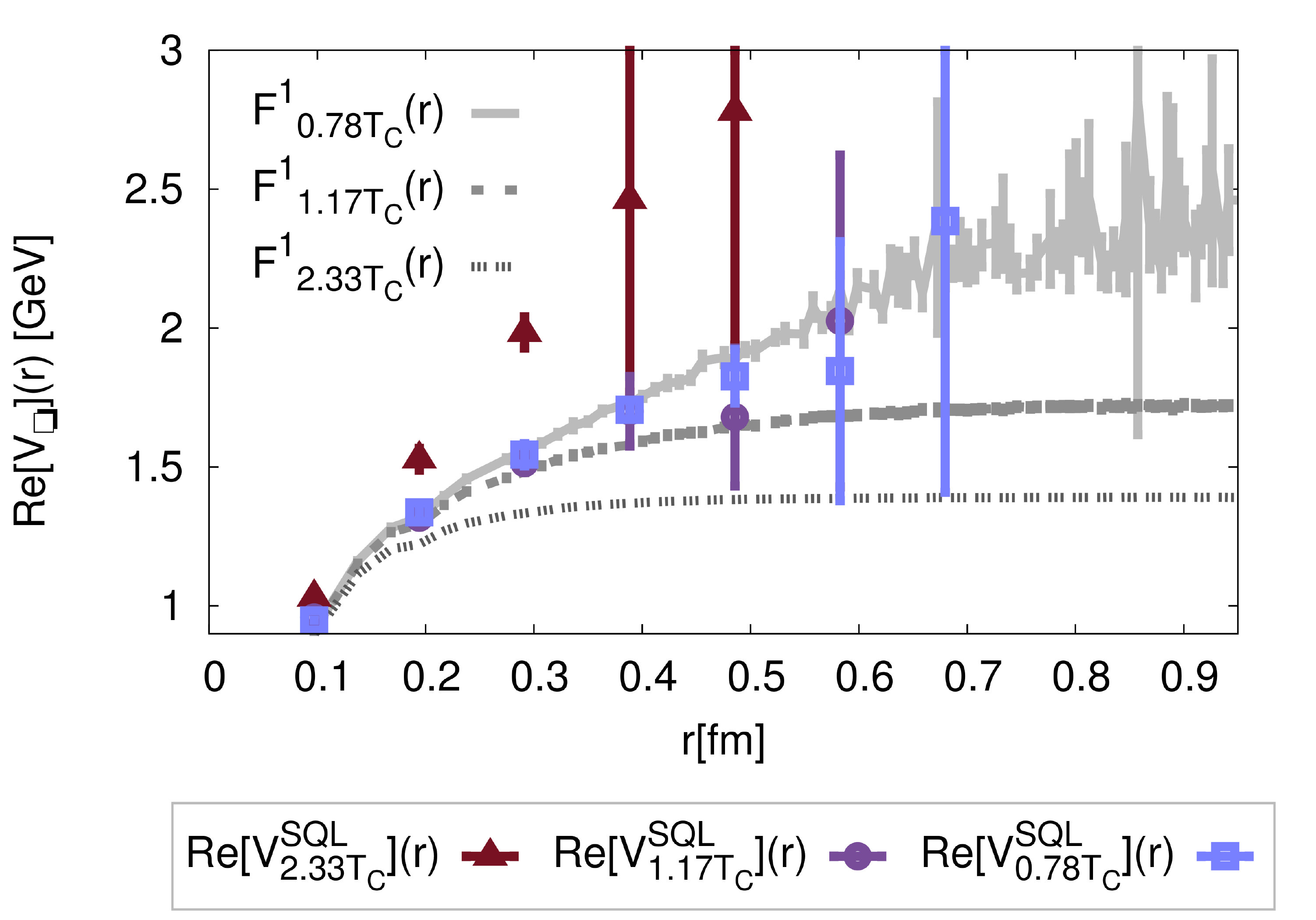}\includegraphics[angle=-90,scale=0.2,clip=true]{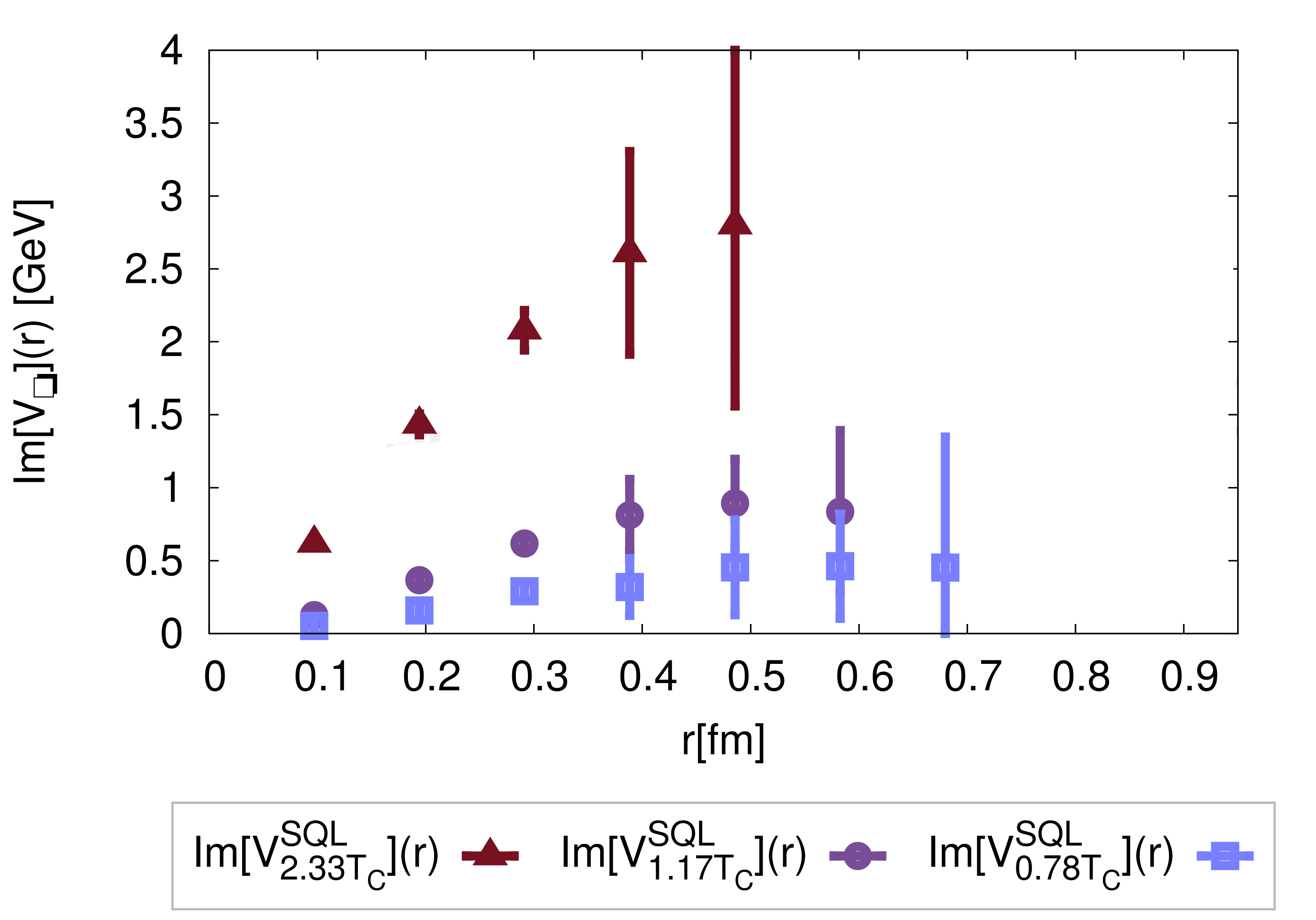}}
\vspace*{8pt}
\caption{The real and imaginary part of the static inter-quark potential in quenched lattice QCD, based on the MEM spectra of Euclidean Wilson loop measurements. We fit the spectrum with Eq.\eqref{Eq:FitShapeFull}, taking into account skewing and up to quadratic background terms (SQL). While as previously indicated, the real part at $T=0.78T_C$ coincides with the color singlet free energies $F^1(r)$, the error bars are still too large to reach a definite conclusion on whether or how the real part at $T>T_C$ exhibits Debye screening. Note that the artificially strong rise from the Breit-Wigner fit disappeared at $T=2.33T_C$. Finite values of ${\rm Im}[V_\square]$ below $T_C$ are most probably due to a finite resolution of the MEM, introduced by statistical uncertainty in the data. In the QGP phase however the curvature of $W_\square(r,\tau)$ leads to finite values. \protect\label{fig2}}
\end{figure}

It is here that we can finally put all pieces of the puzzle together, as sketched in Fig.\ref{fig1}, and present the current state of knowledge on the real and imaginary part of the heavy-quark potential around the phase transition from quenched lattice QCD (Fig.\ref{fig2}). The shown values are based on measurements of the Euclidean Wilson loop\cite{Rothkopf:2011db} in a purely gluonic medium on anisotropic lattices of size $20^3\times12,24,36$ at a bare anisotropy $\xi_b=3.2108$. The choice of $\beta=6.1$ $(a_x=0.097{\rm fm})$ corresponds to the temperatures $T=2.33T_C,1.17T_C$ and $0.78T_C$. After carrying out the MEM on the imaginary time data at different separation distances, the lowest lying positive peak is fitted by the shape introduced in Eq.\eqref{Eq:FitShapeFull} (see also Ref.\cite{Burnier:2012az}) and the corresponding values for ${\rm Re}[V](r)$ and ${\rm Im}[V](r)$ are read off.

\begin{figure}[t!]
\centerline{\includegraphics[scale=0.2,angle=-90,clip=true]{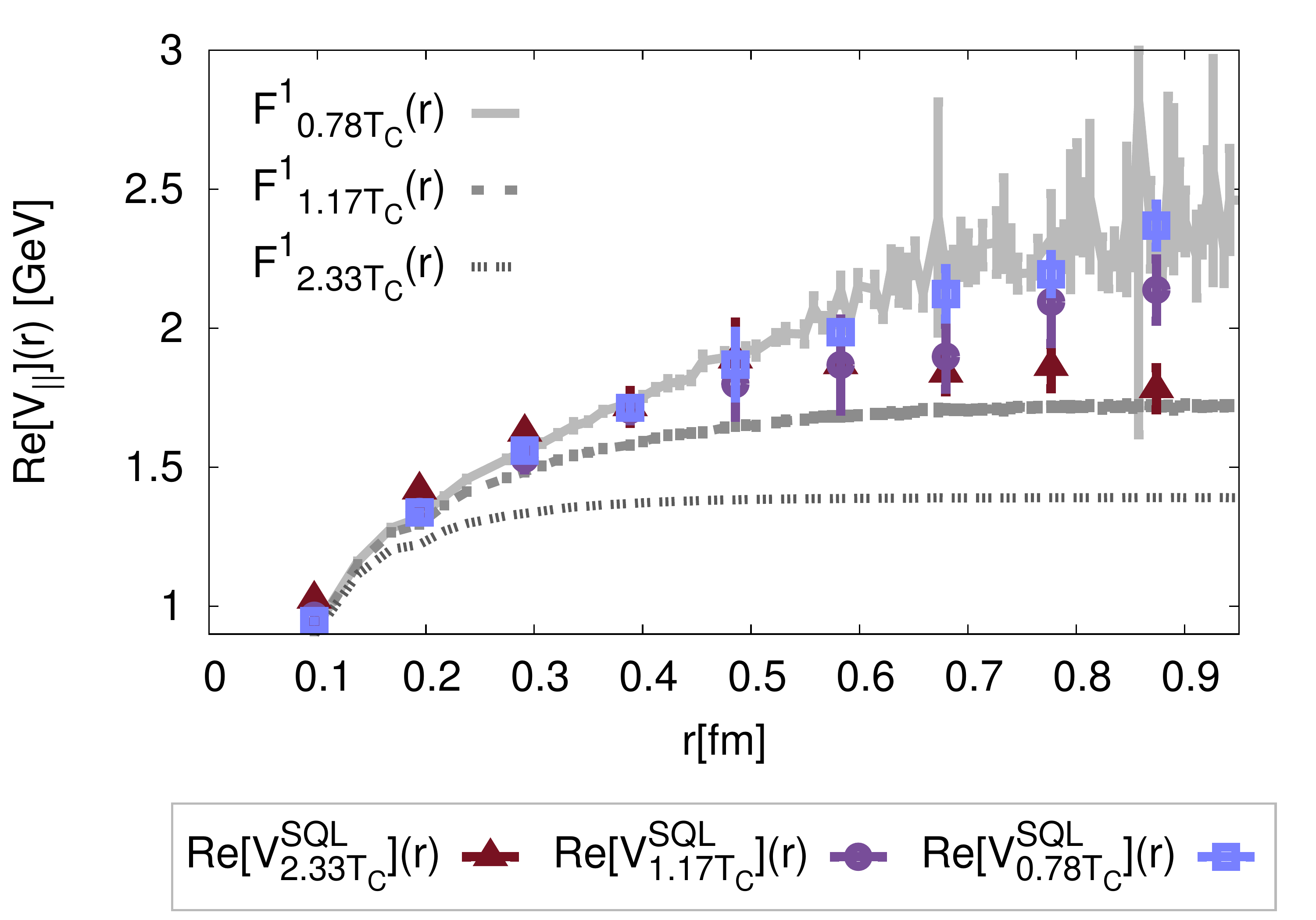}\includegraphics[angle=-90,scale=0.2,clip=true]{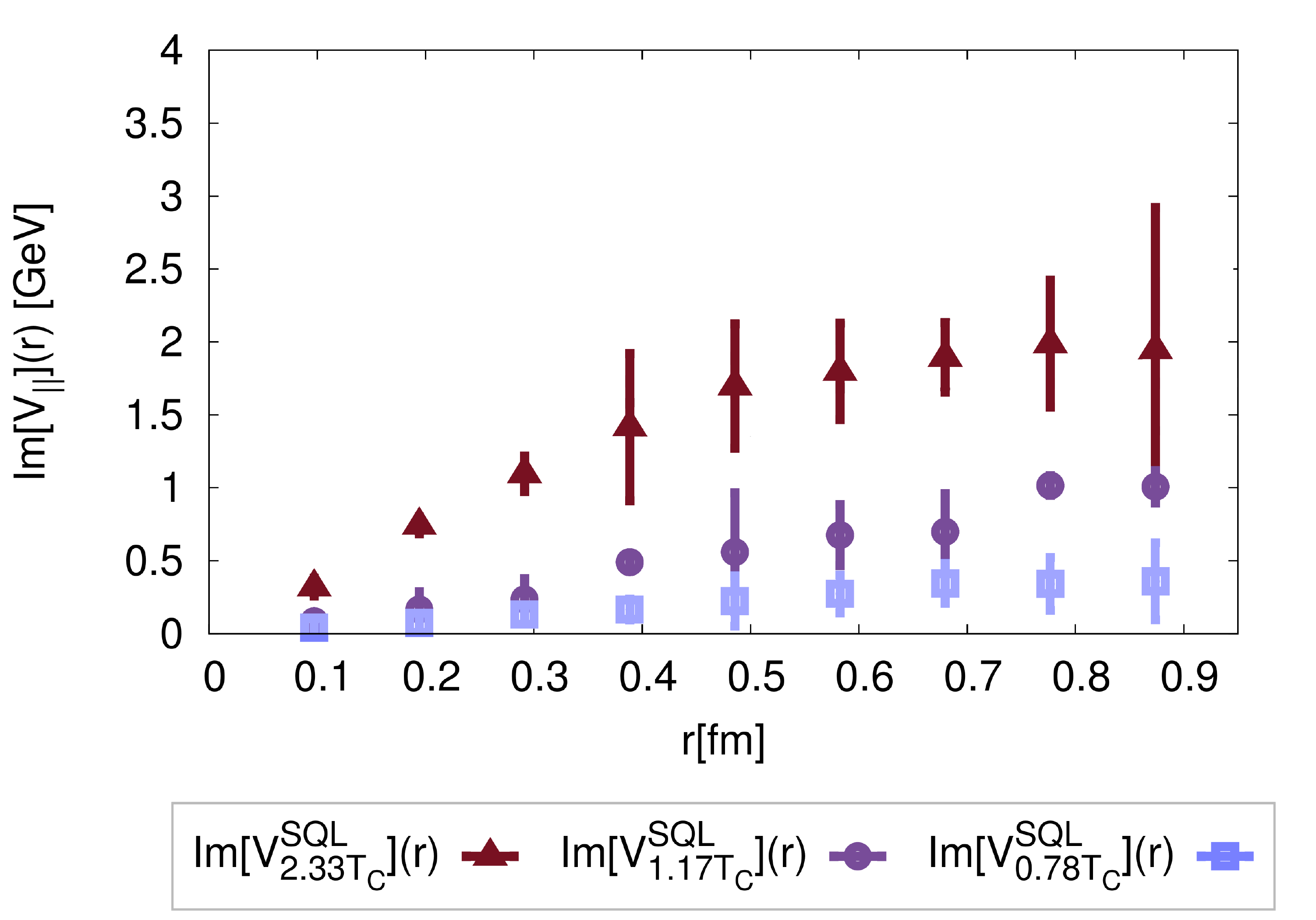}}
\vspace*{8pt}
\caption{The real and imaginary part of the static inter-quark potential in quenched lattice QCD, based on the MEM spectra of Euclidean Wilson line measurements in Coulomb gauge. We fit the spectrum with Eq.\eqref{Eq:FitShapeFull}, taking into account skewing and up to quadratic background terms (SQL). A good signal to noise ratio allows us to see both the linearly rising nature of the real part below $T_C$ as well as a clear sign of screening in the values above $2T_C$. The apparent finiteness of the imaginary part below $T_C$ needs to be attributed to a finite resolution of the MEM, introduced by statistical uncertainty in the data. In the QGP phase at $T=2.33T_C$ the finite imaginary part seems to show a relatively strong rise at $r<0.4{\rm fm}$ while flattening off at larger distances.
\protect\label{fig3}}
\end{figure}

In the hadronic phase, the real part of the potential ${\rm Re}[V_\square]$ agrees with the color singlet free energies $F^1(r)$ within its errorbars. Note that this is not obvious, since the free energies are obtained from the correlator of Wilson lines\footnote{This quantity ensues if the straight spatial Wilson lines $U({\bf x},{\bf y},t)$ are removed from Eq.\eqref{Eq:ForwCorr}} $W_{||}(r,\tau)$ in Coulomb gauge at a single point in time $\tau=\beta$. In the spectral picture (see Fig.\ref{fig1}) this point is not connected to the potential peak at positive frequencies but instead to an exponentially suppressed feature at negative frequencies. 

Above the deconfinement transition the large error bars on the reconstructed values prohibit us from making a conclusive statement on whether or not the potential shows Debye screening, as predicted from HTL calculations. We find that with the the fitting function Eq.\eqref{Eq:FitShapeFull} introduced in Ref.\cite{Burnier:2012az}, the counterintuitively large rise in the real part observed in Ref.\cite{Rothkopf:2011db} at $T=2.33T_C$ is significantly reduced. What is important is that for the imaginary part ${\rm Im}[V_\square]$ we find finite values in the deconfined phase, even after subtracting the values found below $T_C$, as a baseline for the limited resolution of the MEM. Inspecting the Euclidean time data of $W_\square(r,\tau)$ at $T>T_C$, as a cross-check, reveals a significant curvature at intermediate values of $\tau$ which then translates into a width in the spectrum. This gives us confidence that with these extracted values at $T=2.33T_C$ the existence of an the imaginary part in the quark gluon 
plasma is confirmed on the lattice.

If we interpret the Wilson loop as the correlator of two temporal Wilson lines in a spatial axial gauge, we might attempt to redo the extraction of the potential also in Coulomb gauge based on $W_{||}(r,\tau)$. This observable exhibits a much higher signal to noise ratio on the lattice. As shown in Fig.\ref{fig3} it thus allows for a much more accurate determination of the spectral information, which in turn leads to smaller errorbars in both ${\rm Re}[V_{||}]$ and ${\rm Im}[V_{||}]$. The real part appears to move from a confining linear rise, similar to $F^1(r)$, below $T_C$, to a screened form, which however does not coincide with the free energies anymore. The finite values of the imaginary part in the hadronic phase are still to be attributed to an artificial resolution limit in the MEM, coming from the statistical uncertainties of the Euclidean data. The imaginary part at $T=2.33T_C$ on the other hand, which is related to a signal encoded in the Euclidean correlator, data shows a rise at small distances 
$r<0.4{\rm fm}$ before an apparent flattening sets in at larger distances.

The presented results for the static inter-quark potential from lattice QCD tell us that while conceptually a clear path has been established connecting the quantum-chromodynamics of a heavy quark and antiquark in a thermal medium with a potential picture, we need to improve the reliability of the extracted values. One route which needs to be taken is to measure the Wilson loop to much higher accuracy on even larger lattices. As the current calculations are still based on the quenched approximation the multilevel algorithm\cite{Luscher:2001up} might be the appropriate technical tool to use. The MEM itself is known to introduce an artificial width into the reconstructed spectra as its resolution is limited by the amount of uncertainty in the supplied data sets. Hence it needs to be established through mock data analysis how precise we need to measure $W_\square(r,\tau)$ in order to reliably reconstruct the spectral width even below the deconfinement transition. Last but not least, we should extend the 
strategy reviewed here to measurements of the Wilson 
loop in the context of dynamical QCD, where significant effects of deconfined light quarks to the screening of the real part and strength of the imaginary part are expected.

\section{Heavy quarkonium as open quantum system}

The previous sections have shown that the static inter-quark potential in the deconfined phase most certainly exhibits an imaginary part. We might ask how such an apparent non-hermiticity can be interpreted in the context of the in-medium evolution of the heavy-quark bound state. To this end let us remember that we based the definition of the potential on the forward correlator $D^>$ in Eq.\eqref{Eq:ForwCorr} and the presence of a finite ${\rm Im}[V]$ tells us that it is these correlations, which damp away with time. This however does not mean that the constituents of the heavy quarkonium themselves disappear, which they cannot, since annihilation into hard gluons is not taken into account when operating to order $m_Q^{-1}$ in Eq.\eqref{NRQCDLag}. While the overall strength of the correlations is reduced by Debye screening in the presence of deconfined partons, it is scattering with the light quarks and gluons, which actually leads to a decrease in correlations over time. Such a loss of correlation in turn 
manifests itself as an imaginary part in the Schroedinger equation for $ D^>$. This goes so far that after some time, changes in one particle do not influence the state of the other. At this point the notion of a bound state becomes devoid of meaning, the heavy quarkonium has melted.

To formalize this idea of decoherence in the language of quantum mechanics and to see how the imaginary part arises from the thermal fluctuations in the medium surrounding the $Q\bar{Q}$, we turn to a description based on the theory of open quantum systems.(Recent work in this direction can be found in Refs. \cite{Young:2010jq,Borghini:2011yq,Borghini:2011ms,Akamatsu:2011se,Akamatsu:2012vt}) This well established framework provides the conceptual tools to describe the influence of a medium onto a small subsystem, a topic thoroughly investigated in condensed matter theory\cite{Breuer:2002pc}. Assume that both the medium and the $Q\bar{Q}$ can be described quantum mechanically, so that the overall Hamiltonian $H_{\rm full}=H^\dagger_{\rm full}$ is hermitian 
\begin{align}
 H_{\rm full}=H_{\rm med}\otimes I_{Q\bar{Q}}+ I_{\rm med}\otimes H_{Q\bar{Q}}+H_{\rm int}, \quad \frac{d}{dt}\sigma(t)=-i[H_{\rm full},\sigma(t)],
\end{align}
i.e. states evolve unitarily and the density matrix of states $\sigma(t)$ follows the von Neumann equation. If we now wish to describe the system solely in terms of the heavy $Q$ and $\bar{Q}$, as we have done in our attempt to derive an effective Schr\"odinger equation, we have to trace out all other degrees of freedom in the system. Their influence on the evolution of the subsystem manifests itself in the appearance of a stochastic element, such as noise, both in the master equation of the density matrix\cite{Lindblad:1975ef} as well as in the evolution equation of the wavefunction.

Since a Schr\"odinger equation does not possess a notion of thermal fluctuations, one necessarily goes over to an ensemble of wavefunctions $\Psi_{Q\bar{Q}}$ so that the corresponding density matrix of states in the subsystem can be expressed as
\begin{align}
 \sigma_{Q\bar{Q}}(t,{\bf r}, {\bf r}')={\rm Tr}_{\rm med}\Big[\sigma(t,{\bf r},{\bf r}')\Big]=\langle \Psi_{Q\bar{Q}}({\bf r},t)\Psi_{Q\bar{Q}}^*({\bf r}',t)\rangle \label{Eq:QQDensMat}.
\end{align}
By definition, decoherence in this context represents the phenomenon that the interactions with the surroundings select a certain basis of states in the $Q\bar{Q}$ system in which the density matrix $\sigma_{Q\bar{Q}}$ becomes diagonal after the decoherence time $t_{\rm dc}$ has passed. To make as close contact as possible with the potential extracted from lattice QCD, we will however study the influence of the interaction with the thermal medium directly on the level of the wavefunction.

\begin{figure}[t!]
\hspace{-0.75cm}
\centerline{\includegraphics[scale=0.55,clip=true, trim=4.5cm 5.7cm 7cm 8.0cm]{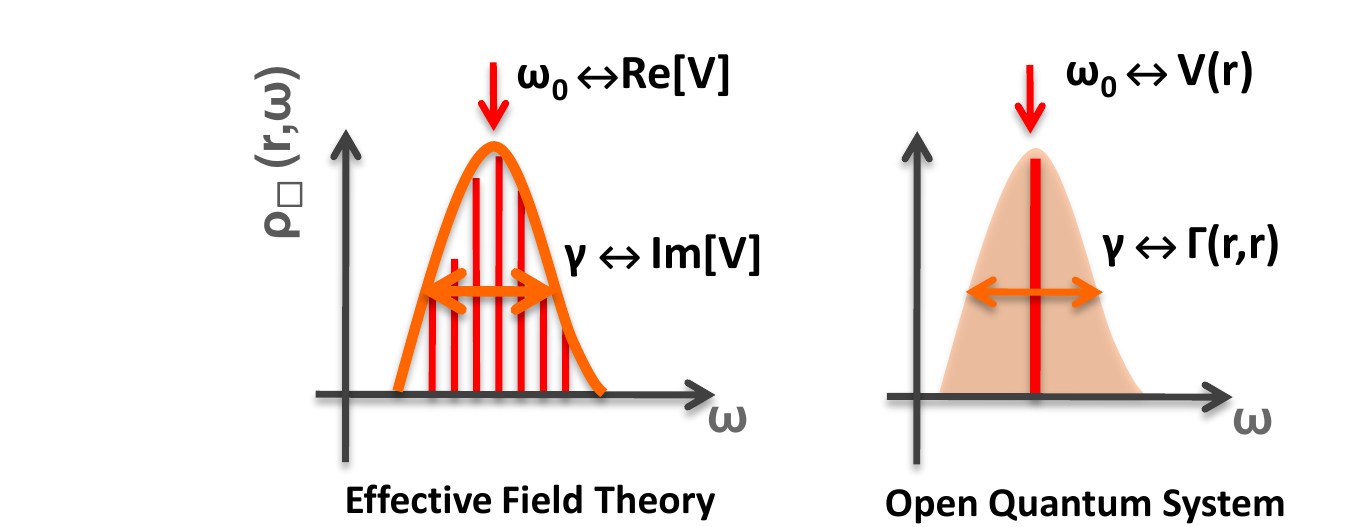} \raisebox{0.35\height}{\includegraphics[scale=0.5]{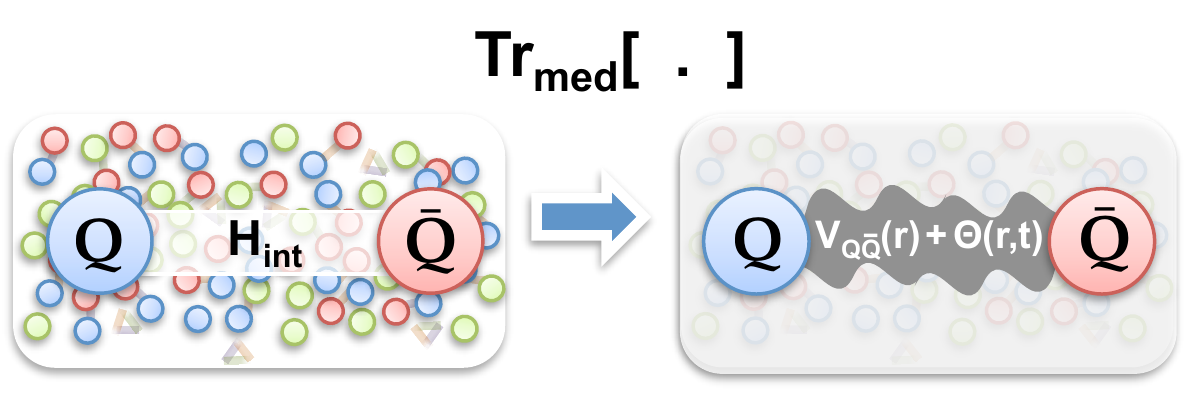}}}
\vspace*{8pt}
\caption{(left) Reinterpreting the spectral features in the Wilson loop spectrum. Instead of assigning the width to an imaginary part of the potential as done in the EFT approach, the open quantum systems approach works with a purely real potential which is perturbed by the thermal medium. It is the strength of the thermal noise $\Gamma(r,r)$ which is then characterized by the width. (right) Schematic view of the difference between a description on the level of the full system with a hermitian Hamiltonian $H=H_{Q\bar{Q}}+H_{\rm med}+H_{\rm int}$ and a stochastic potential description of the $Q\bar{Q}$ system after tracing out the medium degrees of freedom.\protect\label{fig4}}
\end{figure}

In Sec.\ref{Sec:QMPI} we connected the spectral features of the Wilson loop to a complex potential. Here we take a different route\cite{Akamatsu:2011se} as indicated in Fig.\ref{fig4}. The intuitive idea is that the thermal fluctuations, i.e. light quarks and gluons in the QGP, will perturb the potential acting between the heavy $Q$ and $\bar{Q}$ at each step in the time evolution. The average of this purely real potential $V_{Q\bar{Q}}(r)$ corresponds to what was previously called ${\rm Re}[V]$, while its variance is to be connected to the width of the spectral features. Based on this paradigm, let us set out to construct a fully unitary time evolution operator\cite{Akamatsu:2011se} for each microscopic realization of the wavefunction $\Psi_{Q\bar{Q}}$ in the ensemble
\begin{align}
\Psi_{Q\bar{Q}}({\mathbf r},t)={\cal T} {\rm exp}\Bigg[ -i\int_0^t\;ds\;\Big\{ -\frac{\nabla^2}{m_Q}+2m_Q+ V_{Q\bar{Q}}({\mathbf r})+\Theta({\mathbf r},s)\Big\}\Bigg]\Psi_{Q\bar{Q}}({\mathbf r},0)\label{Eq:StochTimeEvol}.
\end{align}
For the sake of simplicity we restrict ourselves to a Markovian noise term $\langle\Theta({\mathbf r},t)\rangle=0$, which however carries a non-trivial spatial correlation structure $\langle \Theta({\mathbf r},t)\Theta({\mathbf r',t'})\rangle=\frac{1}{\Delta t}\delta_{t,t'} \Gamma({\mathbf r},{\mathbf r}')$ characterizing the thermal properties of the QGP medium. An equation of motion for the heavy quarkonium is obtained once the operator of Eq.\eqref{Eq:StochTimeEvol} is expanded following the rules of stochastic differential Ito calculus 
\begin{align}
\hspace{-0.1cm} i\frac{d}{dt}\Psi_{Q\bar{Q}}({\mathbf r},t)=\Big(-\frac{\nabla^2}{m_Q}+2m_Q+V_{Q\bar{Q}}({\mathbf r})+\Theta({\mathbf r},t)-i\frac{\Delta t}{2}\Theta^2({\mathbf r},t)\Big)\Psi_{Q\bar{Q}}({\mathbf r},t)\label{Eq:StochSchroed}.
\end{align}
At this point we are in possession of a stochastic Schr\"odinger equation that preserves the norm of each individual wavefunction of the ensemble while introducing decoherence in the physical heavy quarkonium state, as seen from taking the average
\begin{align}
 i\frac{d}{dt}\langle\Psi_{Q\bar{Q}}({\mathbf r},t)\rangle=\Big(-\frac{\nabla^2}{m_Q}+2m_Q+V_{Q\bar{Q}}({\mathbf r})-\frac{i}{2} \Gamma({\mathbf r},{\mathbf r})\Big)\langle\Psi_{Q\bar{Q}}({\mathbf r},t)\rangle\label{Eq:StochAvgSchroedinger}.
\end{align}
Thus within this open quantum systems approach, the imaginary part of the EFT approach emerges naturally from the diagonal correlations of the thermal fluctuations. On the other hand, extracting the imaginary part of the EFT potential from lattice QCD allows us to pinpoint parts of the noise structure of the QGP.

To describe the suppression of heavy quarkonia, we have to state first what it is that we measure in experiment. Naively speaking, when heavy quarkonium is created in the partonic stage of a collision, it appears as eigenstate $\phi_n({\bf r})$ of a vacuum Hamilton operator $H_{Q\bar{Q}}^{\rm vac}$. After entering the QGP, the stochastic evolution will lead to a reshuffling of states but the particle we finally observe in the detector is still a vacuum eigenstate. Hence we wish to ask how probable it is to find such a vacuum state after a certain time of evolution in the thermal medium, a question, which is answered by the following projection
\begin{align}
 c_{nn}(t)=\int d^3r\,d^3r' \; \phi^*_n({\mathbf r})\langle \Psi_{Q\bar{Q}}({\mathbf r},t) \Psi^*_{Q\bar{Q}}({\mathbf r}',t)\rangle \phi_n({\mathbf r}').
\end{align}
Since this quantity is connected to the density matrix of states introduced in Eq.\eqref{Eq:QQDensMat} it will be susceptible also to the off-diagonal elements in $\Gamma({\mathbf r},{\mathbf r}')$ in contrast to the averaged wavefunction of Eq.\eqref{Eq:StochAvgSchroedinger}. By observing the time evolution of $c_{nn}(t)$ we are confident to learn, within the limitations of the non-relativistic approach, whether and how a heavy quark bound state melts as it evolves in a thermal medium. Conversely by comparison with experimental suppression data, one might attempt to infer properties of the QGP itself.

Obviously we are only at the beginning of a thorough understanding of the open quantum systems nature of heavy quarkonium. One immediate question is e.g. how to extract not only the diagonal but also the off-diagonal components of $\Gamma({\mathbf r},{\mathbf r}')$ from lattice QCD. A more fundamental challenge to the presented approach is the fact that we use a static potential to describe heavy but finite mass quarks in Eq.\eqref{Eq:StochTimeEvol}. As was recently shown in a perturbative rederivation and generalization\cite{Akamatsu:2012vt} of the real-time dynamics inspired by the Feynman Vernon influence functional\cite{Feynman:1963fq}, we are missing a crucial element. 
In fact, dissipation, without which the heavy quarks can never thermalize, must be included to obtain e.g. consistent Ehrenfest relations. Once thermalization has been achieved it will be of interest to compare the resulting distribution of states to the spectral functions from HTL perturbation theory\cite{Burnier:2007qm,Burnier:2008ia,Brambilla:2010vq} and lattice QCD\cite{Asakawa:2003re,Umeda:2002vr,Datta:2003ww,Jakovac:2006sf,Aarts:2007pk,Ding:2010yz,Aarts:2011sm,Ohno:2011zc}.

In addition, a possible route to include explicit color degrees of freedom has been presented\cite{Akamatsu:2012vt}, a feature we require to allow the heavy quarkonium to melt consistently into its constituents. Further studies in this direction should be undertaken as they open the way towards a consistent description of both single heavy quarks and their bound states in the QGP.

\section*{Acknowledgments}
In order to become able to compile this review the author is indebted to the conceptual and technical guidance by T. Hatsuda and S. Sasaki as well as to the insightful discussions and collaboration with Y. Akamatsu and Y. Burnier. A.R. would also like to thank M. Laine and O.Kaczmarek for stimulating discussions and acknowledges partial support by the Swiss National Science Foundation (SNF) under grant 200021-140234.

\end{document}